\documentclass[12pt]{article}
\usepackage{hyperref}
\usepackage{amsmath}
\usepackage{amsthm}
\usepackage{amsfonts}
\usepackage{setspace}
\usepackage[margin=1.0in]{geometry}
\usepackage[numbers]{natbib}
\usepackage{tikz}
\usepackage{multirow}

\usepackage{authblk}
\usepackage{float}
\usepackage{booktabs}
\usepackage{cleveref} 
\usepackage{lscape}
\usepackage{makecell}

\newtheorem{definition}{Definition}

\newtheorem{proposition}{Proposition}
\newtheorem{lemma}{Lemma}

\title{Equitable Discrimination in Survival Prediction:\\ The Maximum Expected C-Index}

\author[1]{
  Felipe Simon\textsuperscript{*}
  }

\author[2]{
 Francisco Pérez-Galarce 
  }
\author[3]{
 Joris van de Klundert
}

\affil[1]{Facultad de Economía y Negocios
  Universidad de Chile}

\affil[2]{Facultad de Ingeniería y Negocios\\
  Universidad de Las Américas}

\affil[3]{Escuela de Negocios\\
  Universidad Adolfo Ibañez}
\affil[ ]{\textsuperscript{*}Corresponding author: \texttt{fsimon@fen.uchile.cl, +569 85017677} }

\begin{document}

\maketitle

\begin{abstract}
    The C-Index measures the discrimination performance of survival prediction models. C-Index scores are often well below the upperbound of 1 that represents perfect prediction and closer to 0.5 as achieved by random prediction. Our first objective is to provide a tighter C-Index upperbound for proportional hazards models. Our second research objective is to measure discrimination performance for subpopulations, also relative to subpopulation specific upperbounds. 
    
    We present the expected C-Index (ECI) as a tight upperbound for proportional hazards models. Moreover, we define the subpopulation C-Index (SUBCI) and a subpopulation specific expected C-Index (SUBECI). The metrics are applied to predict death censored graft survival (DCGF) after deceased donor kidney transplant in the US with a Cox model using standard donor (KDPI), patient (EPTS), and (Class 1) mismatch predictors. With an ECI of 0.75 for 10-year survival, the new upperbound is well below 1.  A C-Index performance around 0.61 or slightly above as commonly reported in literature and replicated in this study therefore closes almost half of the gap between the ECI and the 0.5 threshold. SUBECIs don’t vary significantly from the overall ECI but there are substantial and significant differences among the SUBCIs.  Extending this upperbound and C-Index to subpopulations enables to identify differences in discrimination upperbounds across subpopulations and in prediction model biases. A standard Cox model for DCGF in the US can be ethnically biased.
\end{abstract}

\section{Introduction}

Survival analysis has received considerable attention for a diverse set of populations among which the populations of countries, patient populations, customer segments, and machine collections \citep{cox1972regression, harrell1984regression, dirick2017time, yang2022prognostic}. It has served for descriptive and predictive analytics purposes, such as the prediction of machine breakdown, customer churn, the onset of disease, and death. 

In medical decision making and health policy, survival prediction models are preferred to rely on explicit survival functions \citep{collins2015transparent, wolff2019probast}. A survival function $S_i(t)$ of member $i, i =1,\ldots,m$ of a population $P$ expresses the probability $\mathbb{P}(T_i \geq t)$ that member $i$ experiences a survival ending event at time $T_i \geq t$ and therefore survives at least until time $t$. 

In practical prediction problems, the true survival functions $S_i(t)$ are not known, nor are the corresponding true expected event times $\mathbb{E}[T_i]$. Instead, prediction models rely on estimated survival functions $\hat{S}_i(t)$ and the corresponding expected estimated event time $\mathbb{E}[\hat{T}_i]$ can serve as the predicted survival for population member $i$. 

Depending on the correctness of the model specification, the estimation bias, and the variance of the survival function, the actual $T_i$ may deviate from the predicted $\mathbb{E}[\hat{T}_i]$. Cox's 1972 publication on proportional hazard models has initiated a quest for models to most accurately predict survival \citep{cox1972regression}. This quest has intensified in recent years with the deployment of a wide variety of machine learning models \citep{wang2019machine, wiegrebe2024deep, van2025comparative}.

The performance of prediction models is evaluated using measures from the three categories of calibration, classification, and discrimination \citep{wolff2019probast, collins2015transparent}. Calibration measures such as the $R^2$ and the (Integrated) Brier score capture the difference between the predicted survival probability and the actual survival at a time $t$ (or over a time interval $(0,t]$) \citep{brier1950verification, graf1999assessment}. In survival analysis, classification measures, such as the Area Under the Receiver Operator Characteristic Curve (AUROC), and confusion matrix based measures such as sensitivity and specificity, assess the correctness of classifying which population members survive until a given time $t$ \citep{green1966signal, zweig1993receiver}.

This research focuses on the third category of prediction performance measures, discrimination. In survival prediction, discrimination relates to the question whether population members with longer predicted survival actually survive longer \citep{steyerberg2010assessing}. Discrimination has been argued to be more important than calibration in survival analysis \citep{harrell1984regression, wolff2019probast}. Poorly calibrated models can be recalibrated but such a fix is not readily available in case of poor discrimination. Discrimination is especially important for decision making in resource-constraint settings in which allocation policies consider a ranking of population members by expected survival as a criterion to determine which population members are served first and which members are served later or not at all. 

An important discrimination measure used in survival analysis is Harrell's Concordance Index, henceforth C-Index or $CI$ \citep{harrell1984regression}. The C-Index expresses the concordance between the order of predicted survival times and the order of actual survival times (see below for a formal definition and assumptions). A C-Index score equal to one indicates that the order of the predicted events coincides completely with the order of actual events and a C-Index of zero indicates the predicted order was the complete opposite of the actual order. For a random order, the C-Index converges to $0.5$ as the population size approaches infinity. \\

While a C-Index value of at least 0.7 has appeared in the literature as a prediction performance requirement to apply a prediction model in practice, many medical studies have reported C-Index values below this threshold \citep{rao2009comprehensive, hartman2023pitfalls, van2025comparative}. Moreover, for these applications, new machine learning models often also struggled to improve on the C-Index score achieved by classical Cox proportional hazard models \citep{bae2020machine, herrmann2021large, van2025comparative}. One may therefore wonder whether better performance is possible in these studies or whether the maximum C-Index attainable is a priori bounded from above. Our first research aim is to identify tighter upper bounds for the C-Index. 

With the rise of machine learning models, prediction biases associated with inequities among subpopulations and especially with smaller - minority - subpopulations are of increasing concern \citep{rajkomar2018ensuring, gianfrancescoetal2018,navarro2021risk}. More so, as such prediction inequities may cause or aggravate health inequities when used in medical decision making and health policy \citep{paulus2020predictably, klundert5effectiveness}. Our second research aim therefore regards variation in C-index scores among different subpopulations. In pursuit of this aim, we formally develop a subpopulation C-Index and corresponding upper bounds \citep{klundert5effectiveness}. 

All bounds and C-Indices will be illustrated on the prediction of graft survival after deceased donor kidney transplantation using data from the UNOS/OPTN registry. This analysis will be further strengthened by providing results in which data imbalances and any resulting discrimination biases are corrected by undersampling.

\section{Methodology}

The methodology section is organized as follows. Subsection \ref{survbasics} present an initial analysis of survival functions and derives a first main proposition on discrimination. Subsection \ref{censored} introduces censoring and extends the main proposition to more general survival models that account for censoring. It forms a stepping stone towards Subsection \ref{survmodels} which derives censored upper bounds for the C-Index that can be expected from prediction models. The final subsection \ref{subpopulations} extends the definitions to the levels of subpopulations and individuals as necessary to address biases and equity.

\subsection{Survival functions with proportional hazards} \label{survbasics}

For each member $i, i = 1,\ldots,m$ of a population $P$, the true survival function $S_i(t)$ is the complement of its true time to event function $F_i(t)$, i.e., $F_i(t) = 1-S_i(t)$. $F_i(t)$ thus is the cumulative distribution function describing the probability $\mathbb{P}(T_i \leq t)$ with corresponding probability density function $F^{\prime}_i(t)$ denoted by $f_i(t)$. The corresponding hazard $h_i(t)$ of having an event at time $t \geq 0$ is defined as the probability density of an event at time $t$ divided by the probability of survival until at least time $t$: $h_i(t) = \frac{f(t)}{S_i(t)} = \frac{f(t)}{1-F_i(t)}$ \citep{leemis2023statistical}. 

Following Cox's original model, the analysis below assumes the true hazard functions $h_i(t)$ can be specified as proportional hazard functions \citep{cox1972regression}. The true hazard functions $h_i(t), i = 1,\ldots,n$ then consist of two components; a member independent baseline hazard function $h_0(t)$ and a time independent hazard rate $h_i > 0$: 
\begin{equation}
         h_i(t) = h_i \times h_0(t)  
    \end{equation}
For $i = 1,\ldots,n$ and $t \geq 0$, the corresponding survival functions are: 
\begin{equation}
         S_i(t) = [S_0(t)]^{h_i},   
\end{equation}
where $S_0(t)$ denotes the baseline survival function corresponding to $h_0(t)$ \citep{leemis2023statistical}. By definition, we therefore also have 
\begin{equation} \label{eq:Ft}
         F_i(t) = 1- [S_0(t)]^{h_i}   
\end{equation}
and
\begin{equation}
         f_i(t) = [S_0(t)]^{h_i-1} \times h_i \times f_0(t).  
\end{equation}

Let us now consider the expected event time 
\begin{equation}
    \begin{split}
        \mathbb{E}[T_i] & = \int_0^\infty  t f_i(t) dt \\
& =  \int_0^\infty t dF_i(t) \\
& =   - \int_0^\infty t dS_i(t)\\
& =   \int_0^\infty S_i(t) dt
            \end{split}
\end{equation}
where the last equality follows from integration by parts (as is well known). Thus, the expected event time $\mathbb{E}[T_i]$ equals the expected survival time of population member $i$ and we use them interchangeably in the remainder. The proportional hazard assumption then yields 
\begin{equation}
    \mathbb{E}[T_i] = \int_0^\infty [S_0(t)]^{h_i} dt.
\end{equation} 

It is not hard to verify that for a given baseline survival function $S_0(t)$, $\mathbb{E}[T_i]$ is continuous and strictly decreasing in the hazard rate $h_i > 0$. Thus, for every $\mathbb{E}[T_i] > \lim_{h \rightarrow \infty} \int_0^\infty [S_0(t)]^{h} dt$, there is a unique $h_i$ such that $\mathbb{E}[T_i] = \int_0^\infty [S_0(t)]^{h_i} dt$ and vice versa. The survival functions for all population members $i, i=1,\ldots,n$ can then be specified either by providing $S_0(t)$ and all ${h_i}$ or by providing $S_0(t)$ and all $\mathbb{E}[T_i]$. 

In the special case of a constant baseline hazard rate $h_0(t) = h$ for some constant $h > 0$, the survival function $S_i(t)$ of member $i, i=1,\ldots,m$ is complementary to the exponential distribution, i.e., $S_i(t) = e^{h_i  \times - h \times t}$ \citep{leemis2023statistical}. We then have \\
\begin{equation}
        \mathbb{E}[T_i]  = \frac{1}{h_i \times h}.
\end{equation}

For other complementary distributions of the survival function, such as the Weibull distribution, the relationship between the expected survival time and the hazard rate may also have a closed form expression. In general, however, the true baseline survival and hazard functions can be non-differentiable and non-parametric and calculating the hazard rates for a given baseline hazard function and event time may require numerical approximation. This applies, for instance, when relying on Kaplan-Meier survival curves or on Nelson-Aalen hazard functions. \\ 

With this initial analysis at hand, let us consider the probability that a population member $i$ 'survives' population member $j$. For this comparison, it is important to bear in mind that for each member $i$, $T_i$ is defined relative to a member-specific 'time of birth' $T_i^0$. From this relative perspective, member $i$ can live longer than member $j$ even though  at a certain absolute point in time member $i$ has already had an event while member $j$ has not.

\begin{proposition} \label{mainprop}
    Let $P$ be a population with members $i=1,\ldots,m$ whose survival functions are specified by a population baseline survival function $S_0(t)$ and expected survival times $\mathbb{E}[T_i]$, or corresponding proportional hazard rates $h_i$, and let $p_{ij} 
    =  \mathbb{P} (T_i > T_j)$, i.e. the probability that member $i$ lives longer than member $j$. Then,
    \begin{equation}
        p_{ij} = \frac{h_j}{h_i + h_j}.\\
    \end{equation}
\end{proposition}

\begin{proof}
The probability that member $j$ has had an event by time $T_i$ equals $F_j(T_i)$. Hence,
\begin{equation}
    \begin{split}
\mathbb{P} (T_i > T_j) &  =  \int_{0}^{+\infty} f_i(t)\mathbb{P}(T_i > T_j | T_i = t) dt \\
& =  \int_{0}^{+\infty}  f_i(t)  F_j(t) dt \\
& =  \int_{0}^{+\infty}  f_i(t) \times ( 1 - S_j(t) ) dt \\
& = [F_i(t)]^{\infty}_0 - \int_{0}^{+\infty}  f_i(t) \times S_j(t) dt \\
& = 1 -  \int_{0}^{+\infty}  f_0(t) \times h_i \times S_0(t)^{h_i-1} \times S_0(t)^{h_j} dt \\
& = 1 - \frac{h_i}{h_i+h_j} \int_{0}^{+\infty}  f_0(t) \times ( h_i + h_j) \times S_0(t)^{h_i+h_j-1}  dt \\
& = 1 - \frac{h_i}{h_i+h_j} [F_{[i+j]}(t)]^{\infty}_0 \hspace{3cm} \label{eq:nontriv} \\
& = 1 - \frac{h_i}{h_i+h_j} \\
& = \frac{h_{j}}{h_{i}+h_{j}} \\
\end{split}
\end{equation}
as required. \end{proof}

In the above, (\ref{eq:nontriv}) follows from (\ref{eq:Ft}) and denotes with $F_{[i+j]}(t)$ the cumulative distribution function for the time to event in case of hazard rate $h_i + h_j$.\\

\subsection{Censoring} \label{censored}

A population member $i$ is said to be censored when there is no survival information for member $i$ for part of the time horizon. As a result of censoring, the event time $T_i$ can remain unknown for one or more censored members $i, i = 1,\ldots,m$. The analysis below takes right-censoring into account, which refers to the case in which for every member $i$ there is a {\it time of censoring} $D_i$ and there is no survival information for member $i$ for the time interval $[ D_i, +\infty )$. Right-censoring naturally occurs when population data are collected until a certain (absolute) moment in time (which still implies differences in the $D_i$ because of differences in 'time of birth'). 

Censoring can be relevant in the analysis of survival data even when survival events beyond a certain point in (absolute) time are known and therefore $T_i > D_i$ may exist. In the remainder, {\it uncensored case analysis} refers to survival analysis for populations in which only members for which $T_i$ is known are included, possibly with $T_i > D_i$.

We refer to \citep{wolff2019probast} for a discussion of censoring in survival modeling and its importance. \\

Let $\delta_i = \mathbb{I}(T_i < D_i)$, i.e., it has value 1 in case the survival event of member $i$ happens before member $i$ is censored (if at all) and zero if member $i$ is censored before a survival ending event occurs. Noting that censoring after $T_i$ is not relevant and that censoring often inhibits a later $T_i$, we define:\\

\begin{definition}
For all members $i, i = 1,\ldots,m$ of a population $P$, the \textbf{time of observation} $\tilde{T}_i = T_i \times \delta_i + (1-\delta_i) \times D_i$. 
\end{definition} 

As data on survival are necessarily bounded by the time of data collection, at least one of $T_i$ and $D_i$ will be observed for each member $i$ and the time of observation $\tilde{T}_i$ is the finite minimum of the time of event $T_i$ and the time of censoring $D_i$.\\ 

The actual event time $T_i$ can be unknown when $\delta_i=0$, causing the corresponding hazard rate $h_i$ to be unknown or even undefined. In such cases, we may, however, define a hazard for the time of observation satisfying:
\begin{equation}
    \mathbb{E}[\tilde{T}_i] = \int_{0}^\infty [S_0]^{\tilde{h}_i}dt.
\end{equation}

The following lemma relates $\tilde{h}_i$ to $h_i$ and will be useful in Subsection \ref{survmodels}. \\

\begin{lemma}
\label{lemma:hdec}
   $\tilde{h}_i \ge h_i$ for every patient $i$.
\end{lemma}
\begin{proof}
    If $\delta_i = 1$ then $\tilde{h}_i = h_i$. Now if $\delta_i = 0$ then $\mathbb{E}[T_i] > \mathbb{E}[\tilde{T}_i]$ and because $\int_0^\infty [S_0(t)]^{x}dt$ is decreasing in $x$ we must have that $\tilde{h}_i > h_i$.
\end{proof}

\subsection{An Upper Bound for the C Index of Survival Prediction Models} \label{survmodels}

In this section, we formally define prediction models and then define the C-Index for prediction models. With these definitions at hand, we show how Proposition \ref{mainprop} implies tight upper bounds for the expected C-Index of any prediction model. \\

As time is a continuous variable, all actual event times $T_i$ can be assumed to be distinct and to differ from all censoring times $D_j$. Thus, member $i$ survives member $j$ whenever $\tilde{T}_i > T_j$ but it is not possible to determine whether member $i$ survives member $j$ whenever $\tilde{T}_i > D_j$. For all $i,j, i \neq j$ we accordingly define $A_{ij} = 1$ if $\tilde{T}_i > T_j$ and $0$ otherwise. For a population $P$ the maximum number of valid pairwise survival comparisons $K$ is therefore defined as $K = \sum_{i=1}^m \sum_{j=1, i \not = j}^m A_{ij}$.

We may observe that at for each pair $i,j$, at most one of $A_{ij}$ and $A_{ji}$ can be nonzero and that exactly one of them will be nonzero for the uncensored case analysis and hence that $K = \frac{1}{2} m (m-1)$ for uncensored case analysis. In case of uncensored case analysis, it will be convenient to define $A_{ij}^{U} = 1$ if $T_i > T_j$ and zero otherwise, i.e., in case $T_j > T_i, i,j = 1,\ldots,m, i \neq j$. \\
  
Let us now turn to prediction models and their C-Index:\\

\begin{definition} 
A {\bf prediction model $\hat{M}$} is a set of survival functions $\hat{S}_i(t)$ for each of the members $i, i = 1,\ldots,m$ of population $P$. This prediction model implies a mapping $\hat{M}_{ij}$ from $(ij) \rightarrow \{0,1\}, 1 \leq i,j \leq m, i \neq j$, where $\hat{M}_{ij} = 1$ if $\mathbb{E}[\hat{T}_i] > \mathbb{E}[\hat{T}_j]$ and 0 otherwise. 
\end{definition} 

For clarity of analysis, we additionally assume that the continuous baseline survival function and hazard rates cause all predicted survival times to be distinct. We leave the details of resolving any issues associated with coinciding (predicted) survival and censoring times as they may occur in practical datasets in which time is discretized to Section \ref{casestudyresults}. \\

\begin{definition}
\label{defCIcens}
Let $\hat{M}$ be a prediction model for population $P$ with members $i=1,\ldots,m$. For $i,j, 1 \leq i,j \leq m, i \neq j$, let $A_{ij} = 1$ if $T_i > T_j$ and 0 otherwise. Then, the {\bf C-Index} of $\hat{M}$ is defined as\\
\begin{equation}
    CI(\hat{M}) = \frac{ \sum_{i=1}^{m} \sum_{j=1, j \neq i}^{m} \hat{M}_{ij} \times A_{ij}}{K}. \label{def:CIndex} \\
\end{equation} 
\end{definition}

\begin{definition} \label{def:ECI}
    The {\bf expected C-Index} $\mathbb{E}[CI(\hat{M})]$ of prediction model $\hat{M}$ is defined as \\
\begin{equation}
\mathbb{E}[CI(\hat{M})] = \frac{ \sum_{i=1}^{m} \sum_{j=1, j \neq i}^{m} \hat{M}_{ij} \times \mathbb{E}[A_{ij}]}{K}. \\ \\
\end{equation}    
\end{definition}

Moreover, for uncensored case analysis, we define:\\

\begin{definition} \label{def:ECCCI}
    In case $T_i < \infty$ for all members $i$ of population $P$,
    The {\bf expected \underline{U}ncensored case C-Index} $\mathbb{E}[CI^U(\hat{M})]$ of prediction model $\hat{M}$ is defined as: \\
\begin{equation}
\mathbb{E}[CI^U(\hat{M})] = 
\frac{\sum_{i=1}^{m} \sum_{j=1, j \neq i}^{m} \hat{M}_{ij} \times \mathbb{E}[A_{ij}^U]}{\frac{1}{2}m (m-1)}. \\
\end{equation}    
\end{definition}

The expected C-Index of the true prediction model is an upper bound for the expected C-Index of any prediction model and the expected complete case C-Index of the true prediction model can provide a tighter bound:

\begin{proposition} \label{prop:max:cens}
For any prediction model $\hat{M}$ \\ 
    \begin{equation}
        \mathbb{E}[{CI}(\hat{M})] \leq \mathbb{E}[{CI}(M)].\\
    \end{equation}
Furthermore,
\begin{equation}
        \mathbb{E}[{CI}(M)]  \leq \mathbb{E}[CI^U(M)] \\
\end{equation}
in uncensored case analysis. \\
\end{proposition}

\begin{proof}

We start by showing the second inequality. Proposition \ref{mainprop} assumes uncensored case analysis and together with definitions \ref{def:ECCCI} and \ref{def:ECI} allows to deduce for the true prediction model $M$ that
\begin{align}
    \mathbb{E}[CI^U(M)] &= \frac{ \sum_{i=1}^{m} \sum_{j=1, j \neq i}^{m} M_{ij} \times \mathbb{E}[A_{ij}^U]} {\frac{1}{2}m (m-1)}  \\
    &= \frac{ \sum_{i=1}^{m} \sum_{j=1, j \neq i}^{m} M_{ij} \times \frac{h_j}{h_j + h_i}} {\frac{1}{2}m (m-1)}\\
    &\ge \frac{ \sum_{i=1}^{m} \sum_{j=1, j \neq i}^{m} M_{ij} \times \frac{h_j}{h_j + \tilde{h_i}}} {\frac{1}{2}m (m-1)}\\
    &=\frac{ \sum_{i=1}^{m} \sum_{j=1, j \neq i}^{m} M_{ij} \times \mathbb{E}[A_{ij}]} {K} = \mathbb{E}[CI(M)]
\end{align}
where the inequality follows from \Cref{lemma:hdec} ($\tilde{h} \ge h$).

We move to the first inequality. Noting that the actual event times $T_i, T_j, D_i, D_j$ and therefore also $\mathbb{E}[\tilde{A}_{ij}]$ are independent of the prediction model, we also have 
\begin{equation}
\begin{split}
    \mathbb{E}[CI(\hat{M})] = \frac{ \sum_{i=1}^{m} \sum_{j=1, j \neq i}^{m} \hat{M}_{ij} \times \mathbb{E}[A_{ij}]} { K} \\
    = \frac{ \sum_{i=1}^{m} \sum_{j=1, j \neq i}^{m} \hat{M}_{ij} \times \frac{h_{j}}{h_{j}+\tilde{h}_i} } { K}. \\
\end{split}
\end{equation} 

Any differences between $\mathbb{E}[CI(M)]$ and $\mathbb{E}[CI(\hat{M})]$ must therefore come from differences in the terms $M_{ij} \times \frac{h_{j}}{\tilde{h}_{i}+h_{j}}$ and $\hat{M}_{ij} \times \frac{h_{j}}{\tilde{h}_{i}+h_{j}}$ and hence require that $M_{ij} \neq \hat{M}_{ij}$. This implies that $M$ predicts member $i$ to live longer than member $j$ and therefore that $\hat{M}$ predicts member $j$ to live longer than member $i$, or vice versa. In this case $M_{ij} = \hat{M}_{ji}$. Moreover, by Definition \ref{def:ECI} a difference can only arise when $M_{ij} = \hat{M}_{ji} = 1$. 
    
By definition of $M$, $M_{ij} = 1$ implies $\frac{h_{j}}{\tilde{h}_{i}+h_{j}} > \frac{1}{2} > \frac{\tilde{h}_{i}}{\tilde{h}_{i}+h_{j}}$. Therefore, $M_{ij} \times \frac{h_{j}}{\tilde{h}_{i}+h_{j}} > \hat{M}_{ji} \times \frac{\tilde{h}_{i}}{\tilde{h}_{i}+h_{j}}$.  \\
\end{proof}

\subsection{The subpopulation C-Index} \label{subpopulations}

We now turn to prediction performance for subpopulations. 
 
\begin{definition} \label{def:SCIndex}
Given a population $P$ with members $i, i = 1,\ldots,m$, a partitioning of $P$ into $L$ subpopulations $P_1, P_2, \ldots, P_L$ and a prediction model $\hat{M}$. 
Then, for prediction model $\hat{M}$ and each  $l=1,\ldots,L$, the {\bf subpopulation C-Index} $SUBCI(l,\hat{M})$ is defined as :\\

\begin{equation}
    SUBCI(l,\hat{M}) = \frac{ \sum_{i \in P_l} \sum_{j \in P, j \neq i} [ \hat{M}_{ij} \times A_{ij} + \hat{M}_{ji} \times A_{ji}] }{  \sum_{i \in P_l} \sum_{j \in P, j \neq i} [ A_{ij} + A_{ji} ] }.   \\ \\
\end{equation} 
\end{definition}

For each member $i$ of subpopulation $P_k$, it considers both $A_{ij}$ and $A_{ji}$, at most one of which can have value 1, to ensure all possible pairwise comparisons for members of $A_{ij}$ are included. It is worth noting that Definition \ref{def:SCIndex} is not a {\it within subpopulation} C-Index that might consider how well members of a subpopulation are ranked relative to each other and therefore considers fewer pairwise event time orders:\\

\begin{equation} 
    SUBCI_{\it within} (k, \hat{M}) = \frac{ \sum_{i \in P_k} \sum_{j \in P_k, j \neq i} \hat{M}_{ij} \times A_{ij}}{  \sum_{i \in P_k} \sum_{j \in P_k, j \neq i} A_{ij}}.  \label{def:WithinSubCIndex} \\ \\
\end{equation} 

The subpopulation C-Index enables analysis of (differences in) prediction performance for subpopulations within the context of the population at large. The definition matches a context in which allocation or treatment decisions are taken based on expected survival in relation to the population at large and therefore regardless of subpopulation membership. Moreover, letting $K_l = \sum_{i \in P_k} \sum_{j \in P, j \neq i} [ A_{ij} + A_{ji} ]$ for $l=1,\ldots,L$, the weighted sum
\begin{eqnarray*}
\sum_{l=1}^{L} \frac{| K_l |}{2K} \times SUBCI(l,\hat{M}) & = & \frac{1}{2K} \sum_{l=1}^{L} \sum_{i \in P_l} \sum_{j \in P, j \neq i} [ \hat{M}_{ij} \times A_{ij} + \hat{M}_{ji} \times A_{ji}]\\
  & =  & \sum_{i=1}^{m} \sum_{j=1, j \neq i}^{m} \frac{ [ \hat{M}_{ij} \times A_{ij} ] }{K} \\
  & = & CI(\hat{M}). \\ \\
\end{eqnarray*}

We may note that, for $l=1,\ldots,L$, differences among the $SUBCI(l,\hat{M})$ reflect inequalities in the distributions of estimated survival times among subpopulations within the context of the general population. For example (and in the absence of censoring), when the $\mathbb{E}[\hat{T}_i]$ of members $i$ of subpopulation $l$ are relatively close to the overall mean expected survival time, the $p_{ij}$'s are relatively close to $\frac{1}{2}$ and $\mathbb{E}(SUBCI(l,\hat{M}))$ will be relatively low. Members of subpopulation $P_l$ are harder to rank. This will also and especially be reflected in the $\mathbb{E}(SUBCI_{\it within}(k,\hat{M}))$. 

When the variance in survival times of members of a subpopulation $P_l$ is not different from the variance in survival times for the entire population $P$, but the average survival times tend to be shorter (or longer), the $p_{ij}$ and therefore $\mathbb{E}(SUBCI(l,\hat{M}))$ may be relatively high, without affecting $\mathbb{E}(SUBCI_{\it within}(l,\hat{M}))$. Estimated models $\hat{M}$ can be biased and therefore unequally affect $\hat{p}_{ij}$'s and $\mathbb{E}(SUBCI(l,\hat{M}))$'s. These biases might result in inequities in access to allocation and treatment outcomes when expected survival times co-determine allocation and treatment decisions. 

\section{Methods for Practical Application to Kidney Transplant Survival}

When considering the (expected) C-Indices of prediction models in practice, it is important to distinguish two different settings. The first setting is a truly prospective setting in which predictions about (the order) of survival have to be made before time of observation. A more general version arises when only requiring that predictions have to be made before being informed of times of observation. We refer to the latter as a surrogate prospective setting. The prediction problems in both these settings are considered prospective problems in the remainder.

\begin{definition}
    The data set $D^{pro}$ of a \textbf{ prospective prediction problem} consists of
    \begin{enumerate}
        \item Population $P^{pro}$, with members $i,i,=1,\dots,m$
        \item a partitioning of $P^{pro}$ into $K$ subpopulations, $P_1,\dots,P_K$
        \item for each member $i$, a vector $z_i$ of predictors (or parameter) values $(z_1,\ldots,z_m)^T$,
        \item a specification of the function $g(z_i)$ describing the relationship between the predictor vectors $z_i$'s and the proportional hazard ratios $h_i$, i.e. $h_i = g(z_i)$
        \item a baseline survival function $S_0(t)$,\\
    \end{enumerate}
\end{definition}

Of course, the C-Index performance of a prospective prediction model can only be determined after actual times of survival and censoring are revealed. Moreover, a {\bf retrospective setting} in which actual event times are revealed is required to train - or estimate - a prediction model.

\begin{definition}
    The data set $D^{ret}$ of a {\bf retrospective prediction problem} consists of
    \begin{enumerate}
        \item Population $P^{ret}$, with members $i,i,=1,\dots,m$
        \item a partitioning of $P^{ret}$ into $K$ subpopulations, $P_1,\ldots,P_K$, and
        \item for each member $i$, a vector $z_i$ of predictor (or parameter) values $(z_1,\ldots,z_m)^T$, 
        \item for each member $i$, a transplant time $X_i$ and an actual event time $A_i$ or a time of right censoring (e.g. $t^*$).
    \end{enumerate}
\end{definition}

In practice, models are often specified and estimated using historic data, i.e. on an instance $D^{ret}$, and subsequently applied to make predictions about survival for a population forming a prospective instance $D^{pro}$. A truly prospective prediction problem, for instance, naturally arises when fitting survival models for graft survival after kidney transplant using retrospective data and applying the model to make survival predictions to inform allocation or treatment decisions for future transplants. \\

Evaluating the C-Index performance of a model fit on historic data $D^{ret}$ on future data $D^{pro}$ is a form of so-called temporal validation. Perhaps because of the longitudinal data requirements  - average graft survival after kidney transplantation is in the order of ten years - studies reporting temporal validation of truly prospective prediction problems are scarce in the scientific literature.

An alternative study design adopts a temporal validation approach to a surrogate prospective study design. It splits a retrospective data set into data before a certain point in time $t^*$, defining a data set $D^{ret}$ and a data set $D^{pro}$ with data post $t^*$. After estimating a model on $D^{ret}$, the performance can then be immediately evaluated on $D^{pro}$, enabling temporal validation without delay.

Both forms of temporal validation assume that $D^{ret}$ and $D^{pro}$ are representative samples from a larger population and that the survival functions $S_i(t)$ and the hazard rates $h_i$ estimated for $D^{ret}$ have validity for $D^{pro}$. This assumption may loose validity when the context changes over time, e.g. because allocation systems change or when innovations improve treatment effectiveness. In such cases, poor temporal discrimination performance in the form of a low C-Index score for $D^{pro}$ may be caused by temporal changes in $S_i(t)$ and $g(z_i)$ regardless of model fit for $D^{ret}$. To assess the discrimination performance of a prediction model, one may therefore prefer to split a single data set $D$ randomly into two samples and use one of these samples, $D^{ret}$ in our terminology, to estimate a model and the other as the test data set $D^{pro}$. This can even be repeated multiple times to achieve multi-fold cross-validation. 

Such random split study designs provide better insight into prediction performance for the population from which $D^{ret}$ was sampled but leave the practically relevant validity of the resulting C-Indices for later or future data unaddressed. Indeed, the C-index of such retrospective prediction problems without temporal validation might be viewed as a measure of descriptive performance rather than of predictive performance. 

One may argue that the notion of the expected C-Index becomes less relevant for such descriptive problems as the actual C-Index can be readily obtained. Nevertheless, the maximum expected C-index can still shed light on the prediction performance relative to an upperbound, without being affected by performance issues associated with temporal trends in survival of hazard rates. To minimize the impact of temporal effects, the case study analysis into upperbounds presented below therefore adopts a multi-fold random split design.\\

In any practical application, the true survival functions are not known, which may inhibit the calculation of the upper bound $\mathbb{E}[CI(M)]$. Instead, survival functions have to be estimated in $D^{ret}$. A baseline survival function $\hat{S}_0(t)$ can be estimated for an a priori specified baseline subpopulation $P_0$ by means of a non-parametric maximum likelihood estimator (NPMLE). In the absence of censoring and in the case of independent censoring, the Kaplan-Meier function for $P_0$ maximizes the NMPLE as well as the empirical likelihood \citep{li2005empirical, andersen2012statistical}. Moreover,  it is an unbiased estimator, yielding that $\hat{S}_0(t)$ converges to $S_0(t)$, in the absence of censoring or in the case of independent censoring and $m_0 \rightarrow \infty$ \citep{zhou1988two, stute1994bias}. 

When $D^{ret}$ specifies the predictor values $z_i, i =1,\ldots.m$, the parametric Breslow estimator maximizes the empirical likelihood and is therefore also an unbiased estimator \citep{li2005empirical}. The Breslow estimator can be calculated without specifying a baseline subpopulation $P_0$. Thus, for large enough $m$, unbiased empirical baseline functions $\hat{S}_0(t) = S_0(t)$ can be calculated for retrospective prediction problems, as is adopted in the definition of retrospective prediction problems above. These baseline functions can then feed into an accompanying prospective prediction problem for which members are sampled from the same population.\\

After the transfer of an unbiased baseline survival function $S_0(t)$ from $D^{ret}$ to $D^{pro}$, the above leaves the differences in the estimated versus the true hazard rates as the only unavoidable cause for strict inequality in Proposition \ref{prop:max:cens}. Now, if both $D^{ret}$ to $D^{pro}$ are large enough random samples from the same original data set, any maximum expected C-Index calculated for $D^{ret}$ also applies to $D^{pro}$. For $D^{ret}$, we can calculate unbiased hazard rates ${h}^*_i$ satisfying $\mathbb{E}[T^*_i]  = T_i, i=1,\ldots,m$. Being unbiased, the ${h}^*_i$ can be used to form a prediction model $M^*$ with survival functions $S^*_i(t) = [S_0(t)]^{h^*_i}$. 
Now, because $M_{ij}^* = 1$ if $A_{ij} = 1$ for every pair $i,j$ in $P^{ret}$, $CI(M^*)=1$ in $D^{ret}$. Moreover, $\mathbb{E}[CI(M^*)]$ is an upper bound for the $\mathbb{E}[CI(\hat{M})]$ attainable by any prediction model $\hat{M}$ on $D^{ret}$ and $D^{pro}$. 

Parametric models $\hat{M}$ can use the vectors vector $z_i$ to estimate proportional hazard functions $g(z_i)$ for $D^{ret}$ and use $g$ to predict survival times for $D^{pro}$. The resulting $CI(\hat{M})$ and subpopulation C-indices can be compared to the upper bounds $\mathbb{E}[CI(M^*)]$ and $\mathbb{E}[SUBCI(M_k^*)]$ to assess the prediction performance of $\hat{M}$.

Bearing in mind that random predictions yield an expected C-Index of 0.5 and that $\mathbb{E}[CI(M^*)]$ provides an upperbound on the expected C-Index of any model $\hat{M}$, we introduce a discrimination ratio $DR$ to describe how much of the explainable variability in survival is being identified by $\hat{M}$:
\begin{definition}
    For any population $P$ and any prediction model $\hat{M}$, the {\bf discrimination ratio} $DR$ is defined as
    \begin{equation}
        DR(\hat{M},M^*) = \frac{CI(\hat{M}) - 0.5}{\mathbb{E}[CI(M^*)] - 0.5}.
    \end{equation}
    Moreover for any prediction model $\hat{M}$ and for any subpopulation $P_l$ of a population $P$, the {\bf subpopulation discrimination ratio} is defined as 
    \begin{equation}
        SUBDR(l,\hat{M},M^*) = \frac{SUBCI(l,\hat{M}) - 0.5}{\mathbb{E}[SUBCI(l,M^*)] - 0.5} \\ \\
    \end{equation}
     \end{definition}

\section{Case Study} \label{casestudyresults}

This section illustrates the methodology developed using UNOS data on graft survival after kidney transplant from 2002 to 2019. The dataset considers post transplant survival of 184,235 recipients throughout the US, who are 18 years or older and received a first transplant with a kidney from a deceased donor. It presents upperbounds for the overall population and ethnic subpopulations, as well as the C-indices achieved by a Cox proportional hazard model and corresponding discrimination ratios.

Prediction biases among subpopulations can be caused by data imbalances \cite{collins2015transparent, wolff2019probast}. Hence, in addition to estimating models using all data to create $D^{ret}$, we also report on results obtained for a model in which data imbalances are resolved by undersampling in the form of random deletion of members from the larger subpopulations until all subpopulations considered in $D^{ret}$ are of equal size.

Because the continuous parameter time is typically discretized in datasets, survival times of members are not necessarily distinct in the case study. This can be resolved by small perturbations of the survival times of members with identical survival times. More specifically, for a very small $\epsilon > 0$, the perturbed time to event is $T_i + z_i$, where $z_i \sim Unif(0,\epsilon)$. For small $\epsilon$ and large population sizes the effects of these perturbations on the (expected) c-indices are negligible.  

To ensure robustness of the estimated bounds and models, the data sets considered in the experiments reported below were evenly split (50/50) into $D^{ret}$ and $D^{pro}$ 30 times. The tables report averages and 95 percent confidence intervals for the resulting 30 bounds and C-indices. The choice of the half-half split is based on computational experiments of which the details are provided in  \ref{app:coxmodel}. The details of the preprocessing and the Cox model used to estimate $\hat{M}$ are also provided in  \ref{app:coxmodel}. In short, these models include predictors from KDPI, EPTS, and low resolution class 1 HLA mismatches \citep{rao2009comprehensive, friedewald2013kidney, time2012guide, procurement2018guide}. 

\Cref{tab:ecindexunbalanced} presents the results for survival prediction for eight years of kidney transplants, from 2002 to 2009. Each patient is followed up at least ten years for events up to (and including) 2019. This limit is chosen to avoid noise that could have been introduced by the COVID-19 pandemic. The maximum expected C-index is between $0.775$ and $0.791$ which is far from the $0.609-0.621$ achieved by the Cox model. The discrimination ratio for these models is between $37\%$ and $44\%$, leaving considerable room for improvement, even within the proportional hazards assumption.  

In this data set, censoring can occur because of being lost to follow up, patient death, or because of right censoring on December 31, 2019. The latter form of censoring might explain the decrease in the upper bound over time. Patients who received a transplant in 2002 can have $18$ years of follow up, while at most ten years of follow up is possible for patients who received a transplant in 2009. The transplants of earlier years are therefore less likely to be censored and to have a $D_i < T_i$ as a basis for their hazard rate. This results in a higher variance in the times of observation and therefore a higher upper bound. Interestingly, the C-indices achieved by the Cox models don't benefit from the longer survival data collection periods and as a result their discrimination ratio is relative worse in earlier years.


\Cref{tab:ecindex10year} presents a similar analysis in which survival follow up is right censored after ten years for all patients regardless of year of transplant. Ten years is close to the mean graft survival over the sample period \citep{poggio2021long}. For the case of transplants done in 2009, \Cref{tab:ecindexunbalanced} and \Cref{tab:ecindex10year} refers to the same scenario and the differences in follow up years increase by one for every year earlier. Compared to \Cref{tab:ecindexunbalanced}, the $E[CI(M^*)]$ upper bounds of earlier transplant years are considerably lower and the trend now is that the upper bound increases over time. This is in line with the second inequality introduced in \Cref{prop:max:cens} that compares the expected c-index against the uncensored expected c-index. In \Cref{tab:ecindexunbalanced}, given that there is a longer follow-up period, it should be the case that the percentage of right censored cases would decrease and, therefore, should be closer to the uncensored case compared to the experiments in \Cref{tab:ecindex10year}. Alternatively,  this may be explained by improvements in correctness and completeness of prediction and survival data over time and by treatment improvements. The earlier years now yield higher discrimination ratios, which decrease roughly from $52\%$ to $42\%$. 

Overall, the narrow confidence interval show that results are quite robust and that the differences in the C-Index performance attained by the Cox models sometimes are significant and sometimes not. For both Tables the differences (and trends) in upper bounds are almost all significant.

\subsection{Subpopulation results}
In analogy to \Cref{tab:ecindexunbalanced}, \Cref{tab:ecindex:eth} presents the estimated subpopulation C-Index $SUBCI(l, \hat{M})$ and the maximum expected subpopulation C-index $E[SUBCI(l, M^*)]$ for the four largest subpopulations in the UNOS dataset. These subpopulations are the Asian subpopulation, the Black subpopulation, the Hispanic subpopulation, and the White subpopulation. 



The subpopulation C-indices achieved by the Cox model contrast sharply with the consistency in the upperbounds, as the $SUBCI(l, \hat{M})$ of the subpopulations almost all differ significantly with the $CI(\hat{M})$ reported in \Cref{tab:ecindexunbalanced}. Significance of differences is established by two-sided Sign and Mann-Whitney tests, with a p-value of 0.05 \citep{dixon1946statistical, mann1947test}. 

Data imbalance might be a cause of differences between the $SUBCI(l, \hat{M})$ and the $CI(\hat{M})$. The finding that $SUBCI(l, \hat{M})$ of the smallest sub population, i.e., the Asian population is comparable to the subpopulation C-Idex of the largest subpopulation, the White subpopulation, appears to contradict this explanation.  \Cref{tab:ecindex:eth:balanced} nevertheless presents the analysis results obtained for a data set in which the larger populations are undersampled to obtain equally sized subpopulations.  The balanced data set yields improved results for all subpopulations, especially for the Asian and Hispanic/Latino populations. However, the average difference in C-Index between the Asian and Black sub populations has increased. The poor prediction performance and differences in prediction performance across ethnic subpopulations might cause inequities when predicted survival is used as a basis for allocation or treatment decisions, possibly disadvantaging the Black subpopulation.

%

\section{Discussion}

In response to our first research question, the first main contribution is the upper bound for the expected C-Index presented in Proposition \ref{prop:max:cens} under a proportional hazards assumption. It replaces the trivial C-Index upper bound of 1, which cannot be expected to be achieved and therefore misleading unless survival times are deterministic. The expected C-Index may therefore also serve to outdate the folkloric threshold of 0.7 to assess the quality of prediction model \citep{hartman2023pitfalls}. For prediction models that assume proportional hazards, the proposed discrimination ratio can now serve as a more informative metric of discrimination performance and we therefore recommend its use in future studies. For reference, the code used in the case study is available at \url{https://github.com/simo1148/expected-c-index/tree/main}. 

A second main contribution is the definition of the subpopulation C-Index. It addresses the discrimination performance of survival prediction models for subpopulations and enables to identify prediction biases from a discrimination perspective. While subpopulation based classification and calibration have been reported in the literature \citep{hebert2018multicalibration, kartoun2022prediction, pagano2023bias, ganta2024fairness}, the subpopulation C-Index appears to be the first measure that addresses subpopulation biases of prediction models regarding discrimination. Discrimination is especially important as a measure for prediction model performance as there are no measure to repair poor discrimination performance, as opposed to poor calibration performance \citep{harrell1984regression}. Moreover, biases in prediction performance can translate into inequities in organ allocation and treatment in the world's largest organ allocation schemes of UNOS/OPTN and Eurotransplant which are partially based on Cox proportional hazards models \citep{rao2009comprehensive, opelz2007effect}.

These two main contributions are combined to derive discrimination ratios for subpopulations which can reveal whether differences in subpopulation C-Index scores achieved are due to inherent differences in predictability of events among subpopulations as expressed by the upperbounds, or rather should be attributed by disparities in model fit. If the latter is the case, prediction models might need to be adjusted to equitable, i.e. to avoid unnecesary and unjustified inequalities \citep{braveman2006health, braveman2018health, van2022eliminating}, e.g. by resolving data imbalances among subpopulations \citep{wolff2019probast, van2025comparative}. The discrimination ratio is particularly informative in case survival prediction model results co-determine resource allocation and inequities in discrimination power may translate into inequities in allocation. 

The case study presents results obtained from the large UNOS/OPTN kidney transplant survival data from 2002 to 2019. The upperbound for the expected C-Index is always below 0.8 for periodic subsamples from the 2002-2019 UNOS-OPTN dataset and lower for shorter sample periods. For ten year death censored graft survival - which is close to the mean graft survival time - the expected C-Index was around 0.75. Studies regarding death censored kidney transplant survival on UNOS-OPTN data and otherwise, typically consider much shorter time frames of 1, 3 or 5 years and may therefore have lower expected upper bounds for the recipient population (see e.g. \citep{lentine2024optn}). Still, commonly reported C-Index scores of around 0.61 for UNOS-OPTN data \citep{van2025comparative} reflect a discrimination ratio below 0.5. While machine learning models struggle to improve significantly beyond the performance of Cox models \citep{van2025comparative}, well designed machine learning models may be better suited to close the gap towards the expected C-Index attainable for proportional hazards model than Cox models.

The case study results reveal no substantial differences in the maximum expected C-index across subpopulations. By contrast, the C-Index scores of the Cox models found for the four largest ethnic subpopulations distinguished in the UNOS-OPTN data differ significantly and substantially. While the scores improve after undersampling to correct for data imbalance, the differences remain. Thus, the discrimination ratios for the subpopulations by the standard Cox proportional hazards models reflect prediction biases that may translate into allocation disparities. Given the explainability of the Cox proportional hazards mode, future research to advance understanding of the causes of biases it may produce, impact on allocation, and of solutions is called for. We hope the bounds and case study design will also be valid for studies into C-index performance and discrimination ratios for subpopulations for other survival prediction problems in which inequities may arise and need to be addressed.\\

While the presented upper bound can serve as a reference for any survival prediction, it is only an upper bound for models assuming proportional hazards. The proportional hazards assumption is basic and not always valid. Future research in the quest for better survival prediction models may explore the performance of models that advance beyond the proportional hazards model and can therefore be accompanied by more advanced, and higher upper bounds. New bounds can also be proposed for models relying on stratified baselines. 

Our study has reduced the gap between the lower bound of 0.5 obtained by random concordance prediction and the trivial deterministic upper bound of 1. The case study shows that the gap is roughly reduced by one half with a new upper bound around 0.75 for death censored kidney survival after transplant and the discrimination ratio captures the performance of algorithms relative to the new upperbound. The performance gap can also be reduced by presenting lower bounds for (classes of) models \citep{steck2007ranking}. Tightening the interval is a research direction that can provide model performance guarantees and further demystify the reported performances.

Lastly, it is worth noting that our analysis uses the seminal C-Index definition by Harrell \cite{harrell1984regression} and that generalizations and improvement have been developed since \citep{uno2011c}, \cite{antolini2005time}). Thus, future research developing upper bounds for these later C-index definitions forms a natural and worthwhile extension.

\section{Conclusion}
This paper introduced the expected C-Index as a new upper bound for the discrimination performance of proportional hazard survival models that improves upon the simplistic upperbound of 1. We additionally introduce the (expected) subpopulation C-Index to capture differences in survival predictability among subpopulations and an identify prediction biases. The case study on post kidney transplant (death censored) graft survival shows present an upperbound around 0.75 nevertheless indicating that extent models have considerable room for improvement. For standard Cox models, the case study also reveals discrimination disparities across subpopulations which cannot be attributed to differences in upperbounds and therefore reflect model biases. These findings suggest that current models may cause ethnic disparities when used for resource allocation decisions.


\section*{Data Availability Statement}
The data set is an OPTN star file obtained from UNOS under a data sharing agreement. This data set has been requested for research purposes from UNOS via https://optn.transplant.hrsa.gov/data/view-data-reports/request-data/. 

\section{Funding Statement}
All three authors have been financially supported by Fondecyt Regular grant 1230361 of the Agencia Nacional de Investigacion y Dessarrollo de Chile.

\section{Ethical Approval}
This study has received ethical approval from the Comite de Etica de Investigacion, Universidad Adolfo Ibanez,  decision N°27/2023.


\newpage
\bibliographystyle{unsrt}
\bibliography{refs}

\newpage

\section{Tables}

\begin{table}[ht]
\caption{Expected C-Index vs actual C-Index for multiple years of transplants. The numbers in parentheses are $95\%$ confidence intervals.  }
\centering
\begin{tabular}{lccc}
\toprule
TX year & $CI(\hat{M})$ & $E[CI(M^*)]$ & $DR(\hat{M},M^*)$ \\
\midrule
2002 & 0.614 (0.613,0.616) & 0.791 (0.790,0.792) & 39.28\% \\
2003 & 0.609 (0.607,0.610) & 0.789 (0.789,0.790) & 37.54\% \\
2004 & 0.621 (0.619,0.622) & 0.789 (0.788,0.790) & 41.85\% \\
2005 & 0.610 (0.608,0.611) & 0.787 (0.787,0.788) & 38.19\% \\
2006 & 0.620 (0.618,0.621) & 0.786 (0.786,0.787) & 41.77\% \\
2007 & 0.616 (0.614,0.617) & 0.784 (0.783,0.785) & 40.66\% \\
2008 & 0.620 (0.619,0.622) & 0.778 (0.777,0.779) & 43.30\% \\
2009 & 0.617 (0.615,0.618) & 0.775 (0.775,0.776) & 42.39\% \\
\bottomrule
\end{tabular}
\label{tab:ecindexunbalanced}
\end{table}

\begin{table}[ht]
\caption{Expected C-Index vs actual C-Index for multiple years of transplants. The numbers in parenthesis are $95\%$ confidence intervals. We follow up patients for up to $10$ years for an event.}
\centering
\begin{tabular}{lccc}
\toprule
TX year & $CI(\hat{M})$ & $E[CI(M^*)]$ & $DR(\hat{M},M^*)$\\
\midrule
2002 & 0.614 (0.612,0.616) & 0.740 (0.739,0.741) & 47.41\% \\
2003 & 0.611 (0.609,0.613) & 0.739 (0.738,0.740) & 46.56\% \\
2004 & 0.625 (0.623,0.627) & 0.744 (0.742,0.745) & 51.46\% \\
2005 & 0.610 (0.608,0.611) & 0.748 (0.747,0.749) & 44.16\% \\
2006 & 0.620 (0.618,0.621) & 0.752 (0.751,0.753) & 47.56\% \\
2007 & 0.616 (0.615,0.617) & 0.760 (0.759,0.760) & 44.68\% \\
2008 & 0.621 (0.619,0.622) & 0.762 (0.761,0.763) & 46.01\% \\
2009 & 0.618 (0.617,0.620) & 0.776 (0.776,0.777) & 42.83\% \\
\bottomrule
\end{tabular}
\label{tab:ecindex10year}
\end{table}

\begin{table}[H]
\footnotesize
\caption{Expected C-Index vs actual C-Index for multiple years of transplants. The number in parenthesis are $95\%$ confidence intervals. $\S$ and $\dag$ mark the cases where the Sign and Mann-Whitney tests show significant difference between $SUBCI(l,\hat{M})$ and $CI(\hat{M})$, respectively.   }
\centering
\begin{tabular}{lllcc}
\toprule
TX year & Ethnicity $(l)$ &  $SUBCI(l, \hat{M})$ & $E[SUBCI(l, M^*)]$ & $SUBDR(l, \hat{M}, M^*)$  \\
\midrule

\multirow[m]{4}{*}{2002} & Asian & 0.626 (0.622,0.631)\S \dag & 0.798 (0.796,0.800) & 42.42\% \\
 & Black & 0.600 (0.598,0.602)\S \dag & 0.791 (0.789,0.792) & 34.46\% \\
 & Hispanic/Latino & 0.613 (0.611,0.615) & 0.789 (0.788,0.790) & 39.11\% \\
 & White & 0.622 (0.621,0.624)\S\dag & 0.790 (0.789,0.791) & 42.19\% \\
\cline{1-5}
\multirow[m]{4}{*}{2003} & Asian & 0.622 (0.619,0.626)\S\dag & 0.799 (0.797,0.800) & 40.91\% \\
 & Black & 0.594 (0.592,0.596)\S\dag & 0.787 (0.786,0.788) & 32.70\% \\
 & Hispanic/Latino & 0.615 (0.611,0.618)\S\dag & 0.790 (0.789,0.791) & 39.52\% \\
 & White & 0.613 (0.611,0.615)\S\dag & 0.790 (0.789,0.791) & 38.95\% \\
\cline{1-5}
\multirow[m]{4}{*}{2004} & Asian & 0.645 (0.642,0.648)\S\dag & 0.792 (0.791,0.793) & 49.73\% \\
 & Black & 0.601 (0.600,0.603)\S\dag & 0.790 (0.789,0.791) & 34.99\% \\
 & Hispanic/Latino & 0.624 (0.621,0.626)\S & 0.791 (0.789,0.792) & 42.60\% \\
 & White & 0.630 (0.628,0.632)\S\dag & 0.788 (0.787,0.789) & 45.07\% \\
\cline{1-5}
\multirow[m]{4}{*}{2005} & Asian & 0.633 (0.629,0.637)\S\dag & 0.790 (0.789,0.791) & 45.79\% \\
 & Black & 0.591 (0.589,0.593)\S\dag & 0.787 (0.786,0.788) & 31.55\% \\
 & Hispanic/Latino & 0.619 (0.616,0.621)\S\dag & 0.787 (0.786,0.788) & 41.39\% \\
 & White & 0.619 (0.617,0.620)\S\dag & 0.788 (0.787,0.788) & 41.26\% \\
\cline{1-5}
\multirow[m]{4}{*}{2006} & Asian & 0.618 (0.614,0.622) & 0.777 (0.776,0.778) & 42.75\% \\
 & Black & 0.601 (0.599,0.602)\S\dag & 0.786 (0.785,0.787) & 35.18\% \\
 & Hispanic/Latino & 0.627 (0.625,0.630)\S\dag & 0.789 (0.788,0.790) & 44.00\% \\
 & White & 0.632 (0.630,0.633)\S\dag & 0.787 (0.787,0.788) & 45.84\% \\
\cline{1-5}
\multirow[m]{4}{*}{2007} & Asian & 0.637 (0.633,0.641)\S\dag & 0.785 (0.784,0.787) & 48.02\% \\
 & Black & 0.600 (0.598,0.602)\S\dag & 0.784 (0.784,0.785) & 35.12\% \\
 & Hispanic/Latino & 0.603 (0.601,0.606)\S\dag & 0.787 (0.786,0.788) & 36.06\% \\
 & White & 0.627 (0.626,0.629)\S\dag & 0.783 (0.782,0.784) & 44.99\% \\
\cline{1-5}
\multirow[m]{4}{*}{2008} & Asian & 0.620 (0.617,0.623) & 0.779 (0.777,0.780) & 42.93\% \\
 & Black & 0.602 (0.600,0.604)\S\dag & 0.780 (0.780,0.781) & 36.48\% \\
 & Hispanic/Latino & 0.621 (0.619,0.623) & 0.775 (0.774,0.775) & 44.13\% \\
 & White & 0.636 (0.634,0.637)\S\dag & 0.777 (0.777,0.778) & 48.88\% \\
\cline{1-5}
\multirow[m]{4}{*}{2009} & Asian & 0.614 (0.611,0.617) & 0.774 (0.772,0.775) & 41.55\% \\
 & Black & 0.611 (0.609,0.613)\S\dag & 0.775 (0.774,0.775) & 40.47\% \\
 & Hispanic/Latino & 0.618 (0.616,0.620) & 0.778 (0.777,0.778) & 42.38\% \\
 & White & 0.623 (0.622,0.625)\S\dag & 0.776 (0.775,0.777) & 44.71\% \\
\cline{1-5}
\bottomrule
\end{tabular}
\label{tab:ecindex:eth}
\end{table}

\thispagestyle{empty}
\begin{table}[H]
\footnotesize
\caption{Expected C-Index vs actual C-Index for multiple years of transplants. The number in parenthesis are $95\%$ confidence intervals. $\S$ and $\dag$ mark the cases where the Sign and Mann-Whitney tests show significant difference between $SUBCI(l,\hat{M})$ and $CI(\hat{M})$, respectively. In this case we balance the number of patients of each subpopulations before splitting the data in $P^{ret}$ and $P^{pro}$.}
\centering
\begin{tabular}{lllcc}
\toprule
TX year & Ethnicity $(l)$ &  $SUBCI(l, \hat{M})$ & $E[SUBCI(l, M^*)]$ & $SUBDR(l, \hat{M}, M^*)$  \\
\midrule
\multirow[m]{4}{*}{2002} & Asian & 0.648 (0.644,0.653)\S\dag & 0.803 (0.801,0.805) & 49.04\% \\
 & Black & 0.620 (0.615,0.625)\S\dag & 0.798 (0.796,0.801) & 40.16\% \\
 & Hispanic/Latino & 0.617 (0.611,0.623)\S\dag & 0.794 (0.792,0.796) & 39.80\% \\
 & White & 0.652 (0.649,0.656)\S\dag & 0.795 (0.793,0.797) & 51.70\% \\
\cline{1-5}
\multirow[m]{4}{*}{2003} & Asian & 0.629 (0.625,0.633)\S\dag & 0.804 (0.801,0.806) & 42.46\% \\
 & Black & 0.600 (0.596,0.604)\S\dag & 0.788 (0.786,0.790) & 34.70\% \\
 & Hispanic/Latino & 0.634 (0.629,0.639)\S\dag & 0.790 (0.788,0.793) & 46.10\% \\
 & White & 0.607 (0.602,0.613)\S\dag & 0.789 (0.787,0.792) & 37.10\% \\
\cline{1-5}
\multirow[m]{4}{*}{2004} & Asian & 0.669 (0.664,0.673)\S\dag & 0.795 (0.793,0.796) & 57.19\% \\
 & Black & 0.621 (0.617,0.625)\S\dag & 0.796 (0.794,0.798) & 40.83\% \\
 & Hispanic/Latino & 0.632 (0.627,0.636)\S\dag & 0.794 (0.792,0.796) & 44.77\% \\
 & White & 0.648 (0.644,0.651)\S\dag & 0.793 (0.791,0.795) & 50.33\% \\
\cline{1-5}
\multirow[m]{4}{*}{2005} & Asian & 0.653 (0.650,0.656)\S\dag & 0.794 (0.793,0.796) & 51.93\% \\
 & Black & 0.618 (0.616,0.621)\S\dag & 0.792 (0.790,0.794) & 40.56\% \\
 & Hispanic/Latino & 0.647 (0.644,0.649)\S\dag & 0.781 (0.780,0.783) & 52.10\% \\
 & White & 0.636 (0.632,0.639) & 0.791 (0.789,0.793) & 46.64\% \\
\cline{1-5}
\multirow[m]{4}{*}{2006} & Asian & 0.638 (0.633,0.642)\S\dag & 0.779 (0.778,0.781) & 49.32\% \\
 & Black & 0.618 (0.614,0.621)\S\dag & 0.789 (0.788,0.791) & 40.69\% \\
 & Hispanic/Latino & 0.634 (0.631,0.638)\dag & 0.790 (0.788,0.791) & 46.44\% \\
 & White & 0.638 (0.635,0.642)\S\dag & 0.790 (0.789,0.792) & 47.66\% \\
\cline{1-5}
\multirow[m]{4}{*}{2007} & Asian & 0.657 (0.653,0.660)\S\dag & 0.783 (0.782,0.785) & 55.25\% \\
 & Black & 0.607 (0.603,0.610)\S\dag & 0.786 (0.785,0.788) & 37.35\% \\
 & Hispanic/Latino & 0.628 (0.623,0.632)\S\dag & 0.790 (0.788,0.791) & 44.07\% \\
 & White & 0.646 (0.642,0.650)\S\dag & 0.782 (0.780,0.783) & 51.72\% \\
\cline{1-5}
\multirow[m]{4}{*}{2008} & Asian & 0.634 (0.629,0.639)\S\dag & 0.777 (0.775,0.779) & 48.47\% \\
 & Black & 0.599 (0.595,0.603)\S\dag & 0.780 (0.778,0.782) & 35.29\% \\
 & Hispanic/Latino & 0.625 (0.621,0.629) & 0.772 (0.770,0.774) & 45.87\% \\
 & White & 0.630 (0.625,0.635)\S\dag & 0.775 (0.772,0.777) & 47.34\% \\
\cline{1-5}
\multirow[m]{4}{*}{2009} & Asian & 0.623 (0.618,0.628)\S\dag & 0.777 (0.775,0.779) & 44.41\% \\
 & Black & 0.611 (0.608,0.615)\S\dag & 0.780 (0.778,0.782) & 39.72\% \\
 & Hispanic/Latino & 0.626 (0.622,0.630)\S\dag & 0.783 (0.782,0.785) & 44.41\% \\
 & White & 0.617 (0.613,0.621) & 0.785 (0.783,0.786) & 41.03\% \\
\cline{1-5}
\bottomrule
\end{tabular}
\label{tab:ecindex:eth:balanced}
\end{table}

\newpage
\appendix

\section{Experiment details}
\label{app:coxmodel}

\subsection{Preprocessing}
The first step of the preprocessing removes patients and donors with missing predictor values. The second step removes patients whose age is outside of the interval $[0,100]$. The number of transplants considered in each experiment is presented in \Cref{tab:casenumbers}

\begin{table}[H]
\footnotesize
\caption{Number of transplants in each of the experiments discussed in  \Cref{casestudyresults}.}
\centering
\begin{tabular}{lcclcc}
\toprule
TX year               & \# Cases          & \makecell{\# Cases after \\dropping \\ missing values} & Ethnicity       & \makecell{\# Cases used for \\ Tables \ref{tab:ecindexunbalanced}, \ref{tab:ecindex10year}, \ref{tab:ecindex:eth}} & \makecell{\# Cases used for \\ Table \ref{tab:ecindex:eth:balanced}} \\
\midrule
\multirow[m]{5}{*}{2002} & \multirow[m]{5}{*}{10,785} & \multirow[m]{5}{*}{7,907}          & Asian           & 247      &247              \\
                      &                   &                            & Black           &  1,568      &247             \\
                      &                   &                            & Hispanic/Latino &    716     &247           \\
                      &                   &                            & White           &  2,659   &247               \\
                      &                   &                            & Other           &   154  &90               \\
\cline{1-6}
\multirow[m]{5}{*}{2003} & \multirow[m]{5}{*}{10,890} & \multirow[m]{5}{*}{7,989}          & Asian           & 284       &284            \\
                      &                   &                            & Black           &   1,615   &284              \\
                      &                   &                            & Hispanic/Latino &   741     &284            \\
                      &                   &                            & White           &   2,593   &284              \\
                      &                   &                            & Other           &    145    &109            \\
\cline{1-6}
\multirow[m]{5}{*}{2004} & \multirow[m]{5}{*}{11,687} & \multirow[m]{5}{*}{8,238}          & Asian           &   349    &349             \\
                      &                   &                            & Black           &   1,577    &349             \\
                      &                   &                            & Hispanic/Latino &   840     &349            \\
                      &                   &                            & White           &    2,596  &349              \\
                      &                   &                            & Other           &    111     &113           \\
\cline{1-6}
\multirow[m]{5}{*}{2005} & \multirow[m]{5}{*}{12,426} & \multirow[m]{5}{*}{9,207}          & Asian           &   368       &368          \\
                      &                   &                            & Black           &  1,856    &368              \\
                      &                   &                            & Hispanic/Latino &   921    &368             \\
                      &                   &                            & White           &   2,849   &368              \\
                      &                   &                            & Other           &     144   &136            \\
\cline{1-6}
\multirow[m]{5}{*}{2006} & \multirow[m]{5}{*}{13,179} & \multirow[m]{5}{*}{10,059}          & Asian           &    403    &403            \\
                      &                   &                            & Black           & 2,150   &403               \\
                      &                   &                            & Hispanic/Latino &  975    &403              \\
                      &                   &                            & White           &   3,026  &403               \\
                      &                   &                            & Other           &    136  &144              \\
\cline{1-6}
\multirow[m]{5}{*}{2007} & \multirow[m]{5}{*}{13,340} & \multirow[m]{5}{*}{11,150}          & Asian           &   410    &410             \\
                      &                   &                            & Black           &  2,452   &410               \\
                      &                   &                            & Hispanic/Latino &   1,106  &410               \\
                      &                   &                            & White           &  3,363   &410               \\
                      &                   &                            & Other           &    113   &111                \\
\cline{1-6}
\multirow[m]{5}{*}{2008} & \multirow[m]{5}{*}{13,383} & \multirow[m]{5}{*}{11,693}          & Asian           &   409     &409            \\
                      &                   &                            & Black           &  2,545    &409             \\
                      &                   &                            & Hispanic/Latino &   1,202   &409              \\
                      &                   &                            & White           &    3,488   &409             \\
                      &                   &                            & Other           &     109    &145           \\
\cline{1-6}
\multirow[m]{5}{*}{2009} & \multirow[m]{5}{*}{13,026} & \multirow[m]{5}{*}{11,886}          & Asian           &   463    &463             \\
                      &                   &                            & Black           &  2,684    &463              \\
                      &                   &                            & Hispanic/Latino &   1,165  &463               \\
                      &                   &                            & White           &    3,447   &463             \\
                      &                   &                            & Other           &    154    &154    \\
\cline{1-6}
\bottomrule
\end{tabular}
\label{tab:casenumbers}
\end{table}

\newpage 

\subsection{Cox Proportional Hazards Model Parameters}

The model $\hat{M}$ is fitted using predictors based on the Kidney Donor
Profile Index \cite[KDPI;][]{procurement2018guide} and Estimated Post-Transplant Survival Index \cite[EPTS;][]{time2012guide}. To include  compatibility information the total number of HLA mismatch for the loci A,B,DR forms a final predictor. 

\begin{table}[htbp]
    \centering
    
    \label{tab:parameters}
    \begin{tabular}{|l|l|}
        \hline
        \textbf{Predictor} & \textbf{Data Type} \\
        \hline
        Patient's Age & Number \\
        \hline
        Patient's history of diabetes & Categorical \\
        \hline
        Patient's time on dialysis  & Number \\
        \hline
        Patient's Hepatitis C status  & Categorical \\
        \hline
        Patient-Donor ABO match  & Number \\
        \hline
        Patient-Donor HLA missmatch  & Number \\
        \hline
        Donor's Age & Number \\
        \hline
        Donor's Age$>50$ & Categorical \\
        \hline
        Donor's Height & Number \\
        \hline
        Donor's Weight & Number \\
        \hline
        Donor's Age & Number \\
        \hline
        Donor's history of diabetes & Number \\
        \hline
        Donor's history of hypertension & Number \\
        \hline
        Donor's cause of death & Categorical \\
        \hline
        Serum Creatinine at time of transplant & Number \\
        \hline
    \end{tabular}
    \caption{Predictor Specifications}
\end{table}

\newpage

\subsection{50-50 split for Retrospective and Prospective }

\begin{figure}[h]
\includegraphics[width=0.8\textwidth]{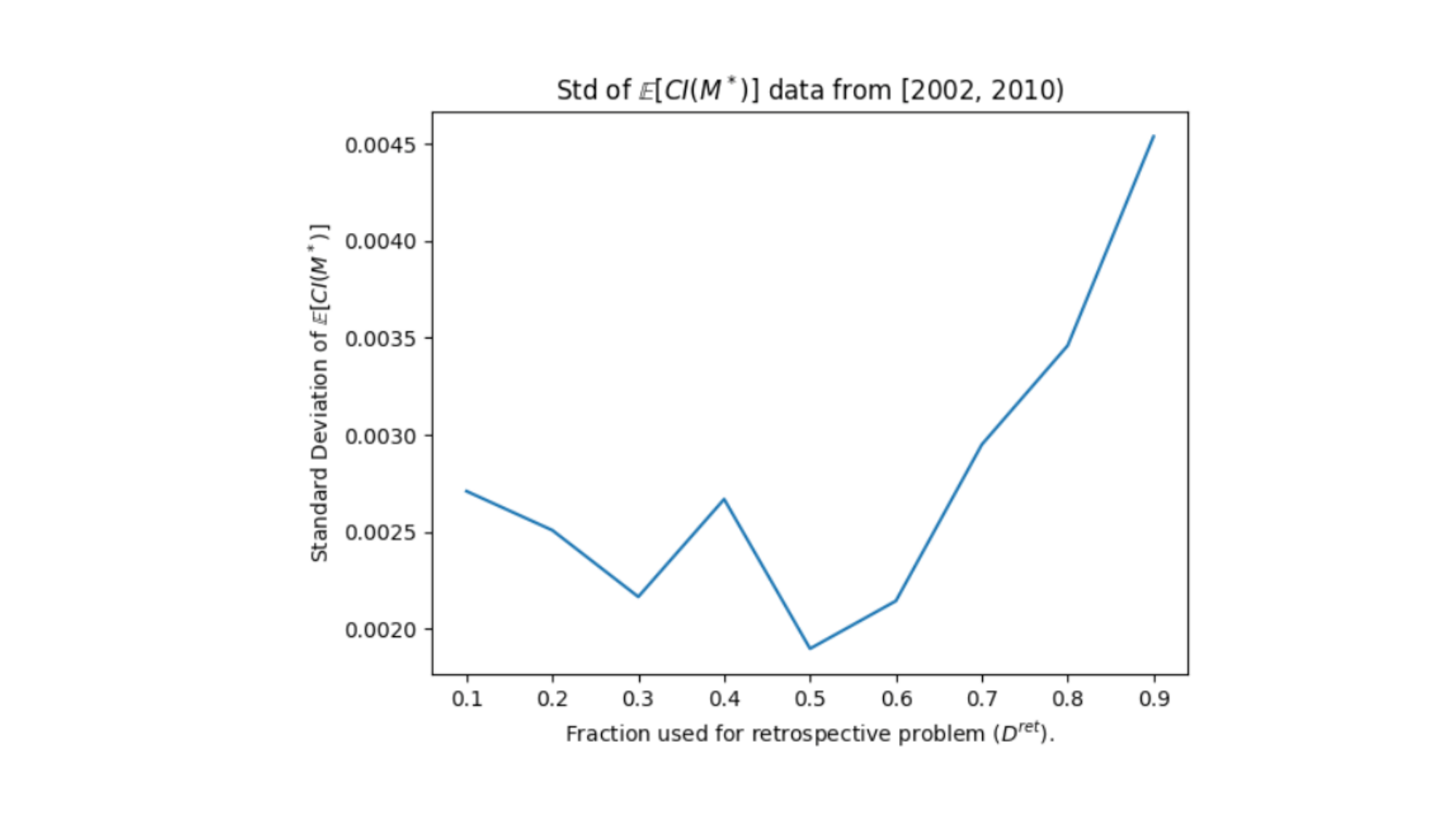}
\centering
\caption{The standard deviation of the expected C-index over $30$ simulations using different proportions for $D^{ret}$ and $D^{pro}$. }
\label{fig:std}
\end{figure}

There is a natural trade-off that arises from the proportion of the population data used to form data set $D^{ret}$ and hence to estimate the baseline cumulative baseline hazard, as well as for the Cox regression. The larger $D^{ret}$, the lower the estimation bias. On the other hand, a smaller $D^{ret}$ allows for a larger $D^{pro}$ and to more accurately assess the resulting C-Index of the estimated Cox models. To quantify this trade-off we calculate the standard deviation of the expected C-Index for different proportions and choose the one that minimizes the variance as this would indicate a balance between unbiased baseline survival estimates and a Cox regression om the one hand and C-index accuracy on the other hand. The results of this experiment are presented in \Cref{fig:std} and show that a 50-50 split minimizes the standard deviation. 

\end{document}